**Title:**

Probabilistic teleportation of a quantum dot spin qubit

**Authors:**

Y. Kojima[1,2*], T. Nakajima[2], A. Noiri[2], J. Yoneda[2,#], T. Otsuka[2,†], K. Takeda[2], S. Li[2], S. D. Bartlett[3], A. Ludwig[4], A. D. Wieck[4], and S. Tarucha[2*]

**Affiliations:**

[1]*Department of Applied Physics, University of Tokyo, 7-3-1 Hongo, Bunkyo-ku, Tokyo 113-8656, Japan*

[2]*RIKEN, Center for Emergent Matter Science (CEMS), Wako-shi, Saitama 351-0198, Japan*

[3]*Centre for Engineered Quantum Systems, University of Sydney, Sydney, New South Wales 2006, Australia*

[4]*Lehrstuhl für Angewandte Festkörperphysik, Ruhr-Universität Bochum, D-44780 Bochum, Germany*

[#]Present address: *School of Electrical Engineering and Telecommunications, University of New South Wales, Sydney, New South Wales 2052, Australia*

[†]Present address: *Research Institute of Electrical Communication, Tohoku University, 2-1-1 Katahira, Aoba-ku, Sendai 980-8577, Japan, Center for Spintronics Research Network, Tohoku University, 2-1-1 Katahira, Aoba-ku, Sendai 980-8577, Japan, and Center for Science and Innovation in Spintronics, Tohoku University, 2-1-1 Katahira, Aoba-ku, Sendai 980-8577, Japan*

[*] e-mail: yohei.kojima@riken.jp or tarucha@riken.jp

**Abstract:**

Electron spins in semiconductor quantum dots have been intensively studied for implementing quantum computation and high-fidelity single- and two-qubit operations have recently been achieved. Quantum teleportation is a three-qubit protocol exploiting quantum entanglement and it serves as an essential primitive for more sophisticated quantum algorithms. Here, we demonstrate a scheme for quantum teleportation based on direct Bell measurement for a single electron spin qubit in a triple quantum dot utilizing the Pauli exclusion principle to create and detect maximally entangled states. The single spin polarization is teleported from the input qubit to the output qubit with a fidelity of 0.91. We find this fidelity is primarily limited by singlet-triplet mixing, which can be improved by optimizing the device parameters. Our results may be extended to quantum algorithms with a larger number of semiconductor spin qubits.

**Introduction:**

An electron spin qubit in semiconductor quantum dot[1] is a promising building block for quantum computing. Recent progress has realized fundamental control on single and two qubits[2–5]. Implementing three-qubit algorithms is a significant step forward, as they can allow for demonstrations of key primitive algorithms such as encoding a single logical qubit[6]. Quantum teleportation[7] (QT) is an attractive instance of three qubit algorithms and has been demonstrated in many physical systems[8–11] because it enables long range quantum communication via quantum repeaters[12] as well as computational models such as gate teleportation[13] and measurement-based control[6]. In quantum-dot spin qubits, however, QT[8,14] has been demonstrated only recently[15], employing a SWAP operation in a Heisenberg spin chain to distribute quantum entanglement in a quadruple quantum dot.

Here, we design a probabilistic QT protocol, where an entangled state is distributed by direct transfer of the qubit with a simple linear ramp pulse in a semiconductor triple quantum dot (TQD) device, and demonstrate the teleportation of a single-electron spin qubit. We employ a sequence of input qubit state preparation, the QT protocol and output qubit state readout by energy-selective tunneling (Fig. 1a). We show that the spin polarization of the input qubit is teleported to that of the output qubit when and only when we have access to the outcome of the Bell measurement. This agrees with the essential property of QT requiring not only quantum entanglement but also classical information about an outcome of the Bell measurement. We analyze possible error sources in our QT process based on input-output relation and find that the teleportation infidelity originates primarily from the leakages of singlet states to and from spin polarized triplet states in the preparation and the measurement of the singlets.

**Results:**

**The QT protocol**

Figure 1b shows the QT protocol in a TQD. The top dot (QD1) hosts an input qubit Q1 and the bottom dot (QD3) hosts an output qubit Q3, while the middle dot (QD2) serves as a transport channel of the ancillary qubit Q2. First, the singlet state $|S_{23}\rangle = \frac{|\uparrow_2\downarrow_3\rangle - |\downarrow_2\uparrow_3\rangle}{\sqrt{2}}$ is created in Q2 and Q3 as the maximally entangled pair by initializing a doubly-occupied spin singlet in QD3 and the electron hosting Q2 is moved to QD2. Then, the electron is attempted to be transferred to QD1. In this process, the Bell measurement of Q1 and Q2 is implemented by singlet-triplet readout using Pauli spin blockade (PSB)[16,17], where the tunneling of the electron hosting Q2 into QD1 coincides with the projection of Q1 and Q2 onto the singlet state $|S_{12}\rangle = \frac{|\uparrow_1\downarrow_2\rangle - |\downarrow_1\uparrow_2\rangle}{\sqrt{2}}$. The Bell measurement only distinguishes the singlet from the other three maximally entangled states, which makes success of the QT stochastic. The state of Q3 becomes the same as Q1 when the outcome of the Bell measurement is singlet. The key ingredients in our QT protocol are coherent separation of $|S_{23}\rangle$ and detection of $|S_{12}\rangle$. Our protocol realizes these steps not with two-qubit gates but via rapid adiabatic passage[16] employing a linear ramp of the detuning energy $\epsilon$, where $\epsilon$ is defined as the energy difference between (2,0,1) and (1,0,2) (see Fig.1d). An advantage of this approach is that the pulse sequence is less complex and the entire operation time is within nanoseconds under appropriate inter-dot tunnel couplings.

**Triple quantum dot device**

A linearly coupled TQD[18,19] is fabricated on a GaAs/AlGaAs heterostructure wafer as shown in Fig. 1c. We apply voltage pulses to the P1 and the P3 gates to rapidly control energy levels of the TQD. A micromagnet fabricated on the wafer surface forms an inhomogeneous local magnetic field and enables addressable electric-dipole spin resonance (EDSR)[3,20] control. The local magnetic field is largest in QD1 followed in order by in QD2 and QD3 (the static external magnetic field is 3.07 T). We detect the electron charge configuration in the TQD by measuring the conductance of the nearby sensor dot as demodulated reflectometry signal ($V_{\text{rf}}$)[21]. Figure 1d shows the charge stability diagram of the TQD around the charge states of $(N_1, N_2, N_3) = (1,1,1)$, $(1,0,2)$ and $(2,0,1)$ used in this work, where $N_i$ denotes the number of electrons in QDi. Note that we tune the inter-dot tunnel couplings so that direct spin-spin interaction between QD1 and QD3 in $(1,1,1)$ is negligible (see Supplementary I).

**Ingredients of the QT protocol**

The fidelity of the entire process of our QT protocol is subject to the tunnel coupling strengths because our approach relies on the mapping between the singlet states and the doubly occupied charge states during detuning ramps. For example, the weak inter-dot tunnel couplings may cause failure in detection of $|S_{12}\rangle$ and separation of $|S_{23}\rangle$ due to slow electron tunneling of Q2 (see Fig. 1b). To confirm the feasibility of $|S_{12}\rangle$ detection, we perform the PSB measurement with QD1 and QD2 (Fig. 2a). Figure 2b shows the histogram of the single-shot PSB signal $V_{\text{rf}}$ measured for QD1 and QD2 after loading a spin-up into QD1 and a random spin into QD2. We use a latched readout technique to enhance the readout visibility, which transfers the spin-blocked $(1,1,1)$ charge state to $(2,1,1)$ before the readout[22,23]. The solid line is a fit using two noise-broadened Gaussian distributions considering the relaxation of triplet states[17]. The state is registered as singlet (triplet) when $V_{\text{rf}}$ is lower (higher) than the threshold voltage, $V_{\text{threshold}}$. Next, we measure singlet-triplet oscillation (ST oscillation) in QD2 and QD3 induced by the local Zeeman field difference[24] $\Delta B_z$ to ensure the creation and separation of $|S_{23}\rangle$ (Fig. 2c). Figure 2d shows the measured singlet probability, $P_S$, as a function of the dwell time $t_{\text{dwell}}$ in $(1,1,1)$. Because the dwell point is far detuned from the $(1,1,1)$-$(1,0,2)$ degeneracy point, the exchange coupling between Q2 and Q3 is suppressed and the observed periodic oscillation of $P_S$ indicates coherently repeated transitions between the singlet and non-polarized triplet[17] (The oscillation visibility is largely limited by readout error arising from the relaxation of non-polarized triplet, which does not contribute to the teleportation infidelity). These results (Fig. 2b and Fig. 2d) show that the device is properly set up to realize coherent separation of $|S_{23}\rangle$ and projection measurement onto $|S_{12}\rangle$.

**Preparation of input qubit and readout of output qubit**

To prepare an input state for the QT protocol, we rotate Q1 using resonantly driven coherent oscillation. We can use a micromagnet-mediated EDSR[20] for the qubit rotation, but here we find that we can manipulate Q1 with higher speed and less decay by resonant transitions between $|\uparrow_1 \Downarrow\rangle$ and $|\downarrow_1 \Uparrow\rangle$ in QD1 and QD2 (resonant SWAP)[25,26] than by EDSR. The double lines arrow ($\Uparrow$ or $\Downarrow$) represents a spin that is temporarily loaded in QD2 to assist the rotation of Q1 and is later discarded to the reservoir. We implement this resonant SWAP by applying a microwave (MW) to the P2 gate after initializing a singlet state in QD1 and subsequently loading $|\uparrow_1 \Downarrow\rangle$ using slow adiabatic passage[27] (Fig. 2e). The resulting two-spin state is $\alpha|\uparrow_1 \Downarrow\rangle + \beta|\downarrow_1 \Uparrow\rangle$ with $|\alpha|^2 + |\beta|^2 = 1$. By emptying QD2, the coherence of Q1 is lost but the probabilities in the up/down basis $|\alpha|^2$ and $|\beta|^2$ are retained. Therefore, $|\alpha|^2$ ($|\beta|^2$) is equal to the spin-up (spin-down) probability of Q1, $P_{\uparrow,\text{in}}$ ($P_{\downarrow,\text{in}}$). To estimate $P_{\uparrow,\text{in}}$, we measure the probability of $|\uparrow_1 \Downarrow\rangle$ ($P_\uparrow^{\text{raw}}$) after a MW burst. The measured probability distribution $P_\uparrow^\text{raw}$ is influenced by the readout infidelities, $1 - f_{\uparrow,\text{in}}$ and $1 - f_{\downarrow,\text{in}}$ for the spin-up and –down state, respectively, and the discrepancy from the actual probability may lead to the underestimation of the performance of our QT protocol. We exclude the readout infidelities as $P_{\uparrow,\text{in}} = \frac{P_\uparrow^\text{raw} + f_{\downarrow,\text{in}} - 1}{f_{\uparrow,\text{in}} + f_{\downarrow,\text{in}} - 1}$ with $f_{\uparrow,\text{in}} = 0.96$ and $f_{\downarrow,\text{in}} = 0.90$ (see Supplementary II). Figure 2f shows $P_{\uparrow,\text{in}}$ as a function of the MW burst time $t_\text{burst}$, indicating that we can vary the spin-up probability of the input state.

The output qubit state Q3 teleported from Q1 is read out by the spin-selective tunneling to the lower reservoir[28]. This readout is performed by pulsing gate voltages near the (2,0,1)-(2,0,0) transition line (marked by a star in Fig. 1d). The tunneling of Q3 to the reservoir occurs only when its spin is down. We estimate the output qubit readout fidelities for spin-up and –down state to be $f_{\uparrow,\text{out}} = 0.83$ and $f_{\downarrow,\text{out}} = 0.50$ from additional experiments (see Supplementary III). $f_{\downarrow,\text{out}}$ is limited by a relatively small tunnel rate to the reservoir compared to the readout time. We estimate the spin-up probability by taking into account those infidelities — similarly to Q1.

**Demonstration of QT**

We now integrate all operations including the preparation of Q1, our QT protocol and the final readout of Q3 in one sequence. Figure 3a shows the spin-up probability of Q3 obtained as a function of $t_\text{burst}$ to drive Q1. The gray squares denote $P_{\uparrow,\text{out}}$, the spin-up probability produced without using the information of the Bell measurement. $P_{\uparrow,\text{out}}$ is independent of $t_\text{burst}$, showing no correlation with the input spin Q1. In contrast, if we extract the data set conditioned on the singlet outcome in the Bell measurement (classical information, CI), we obtain $P_{\uparrow,\text{out}}^\text{CI}$ (see blue circles in Fig. 3a), which reproduces the Rabi oscillation of Q1 (Fig. 2f) well. Here, $V_\text{threshold}$ of the Bell measurement is chosen to take advantage of classical information maximally (see the black dashed line in Fig. 3B), i.e., to maximize the oscillation amplitude $A_\text{out}$ of Q3 spin-up probability. As the accuracy of the classical information is degraded deliberately by raising $V_\text{threshold}$, a monotonic decrease of the amplitude is observed. These agree well with an essential property of the QT, that useful information cannot be extracted only from the local measurement of Q3 and the outcome of the Bell measurement is required to reproduce the original state of Q1. The difference between $P_{\uparrow,\text{out}}$ and $P_{\uparrow,\text{out}}^\text{CI}$ and the similarity between $P_{\uparrow,\text{in}}$ and $P_{\uparrow,\text{out}}^\text{CI}$ are the hallmark of successful teleportation.

**Discussion**

When there are no errors in the singlet preparation and detection, $P_{\uparrow,\text{out}}^\text{CI}$ obtained after our QT protocol should be identical to $P_{\uparrow,\text{in}}$. We plot these two probabilities against each other in Fig.4 (pink triangles). The discrepancy between the two suggests that such errors are indeed not negligible in our experiment. We discuss below the effects of those errors in our QT protocol.

The error in the singlet preparation can be caused during the transition of Q2 from QD3 to QD2. Two-spin states in QD2 and QD3 are generally described by a combination of $|\uparrow_2\downarrow_3\rangle$, $|\downarrow_2\uparrow_3\rangle$, $|\uparrow_2\uparrow_3\rangle$ and $|\downarrow_2\downarrow_3\rangle$. However, the leakage error into $|\downarrow_2\downarrow_3\rangle$ is negligible because this state is energetically separated by the large magnetic field. $|\uparrow_2\uparrow_3\rangle$ is similarly separated but transverse magnetic field difference $\Delta B_x$ may mix $|S_{23}\rangle$ and $|\uparrow_2\uparrow_3\rangle$ at their degenerate point during the ramp from (1,0,2) to (1,1,1). As this transition occurs coherently following the Landau-Zener transition[29,30], the prepared state before the Bell measurement can be described as $|\Psi_{23}\rangle = \gamma|\uparrow_2\downarrow_3\rangle + \delta|\downarrow_2\uparrow_3\rangle + \zeta|\uparrow_2\uparrow_3\rangle$ (1) with $|\gamma|^2 + |\delta|^2 + |\zeta|^2 = 1$. Here, the perfect singlet preparation would lead to $\gamma = \frac{1}{\sqrt{2}}$, $\delta = -\frac{1}{\sqrt{2}}$ and $\zeta = 0$.

The second source of error is in the detection process. This can be decomposed to the state-mapping error and the electrical detection error in the Bell measurement. The latched readout technique employed here helps to suppress the electrical detection error to 0.02 but a large state-mapping error may remain due to the nonideal inter-dot and dot-to-lead tunnel rates[22,31]. To model the error in the detection process, we use the measurement operator $M_{\text{Bell}}$ as

$$M_{\text{Bell}} = F_{S,\text{Bell}}|S_{12}\rangle\langle S_{12}| + (1 - F_{T_0,\text{Bell}})|T_{0,12}\rangle\langle T_{0,12}| + (1 - F_{\Phi,\text{Bell}})(|\Phi_{12}^+\rangle\langle\Phi_{12}^+| + |\Phi_{12}^-\rangle\langle\Phi_{12}^-|), \quad (2)$$

where $1 - F_{S,\text{Bell}}$, $1 - F_{T_0,\text{Bell}}$ and $1 - F_{\Phi,\text{Bell}}$ are the detection errors of the singlet $S$, non-polarized triplet $T_0$ and the other Bell states $\Phi^\pm$ in the partial Bell measurement.

Using Eqs. 1 and 2, we now calculate the final spin-up probabilities of Q3 expected from our model with and without CI, which we denote as $P_{\uparrow,\text{out}}^{\text{CI,model}}$ and $P_{\uparrow,\text{out}}^{\text{model}}$, respectively (see Supplementary IV). The density matrix of the three-qubit state before the Bell measurement $\rho_{123}$ is expressed as $\rho_{123} = \rho_1 \otimes \rho_{23}$, where $\rho_1$ is the density matrix of Q1 and given by $P_{\uparrow,\text{in}}|\uparrow_1\rangle\langle\uparrow_1| + P_{\downarrow,\text{in}}|\downarrow_1\rangle\langle\downarrow_1|$, and $\rho_{23} = |\Psi_{23}\rangle\langle\Psi_{23}|$. The singlet probability $P_{S,\text{Bell}}^{\text{model}}$ in the Bell measurement is then given by

$$P_{S,\text{Bell}}^{\text{model}} = \text{Tr}(\rho_{123}M_{\text{Bell}}^\dagger M_{\text{Bell}}) = F_{S,\text{Bell}}p_S + (1 - F_{T_0,\text{Bell}})p_{T_0} + 2(1 - F_{\Phi,\text{Bell}})p_\Phi, \quad (3)$$

where $p_{S(T_0,\Phi)}$ is the probability of detecting $S$ ($T_0, \Phi$) in the Bell measurement without any detection error. Note that $P_{S,\text{Bell}}^{\text{model}}$ depends on the state of Q1 unless the preparation of singlet is perfect. Using this notation, we obtain $P_{\uparrow,\text{out}}^{\text{CI,model}}$ and $P_{\uparrow,\text{out}}^{\text{model}}$ as follows:

$$P_{\uparrow,\text{out}}^{\text{model}} = \langle\uparrow_3|\text{Tr}_{12}(\rho_{123})|\uparrow_3\rangle = |\delta|^2 + |\zeta|^2 \quad (4),$$

$$P_{\uparrow,\text{out}}^{\text{CI,model}} = \frac{\langle\uparrow_3|M_{\text{Bell}}\rho_{123}M_{\text{Bell}}^\dagger|\uparrow_3\rangle}{\text{Tr}(\rho_{123}M_{\text{Bell}}^\dagger M_{\text{Bell}})}$$

$$= \frac{(F_{S,\text{Bell}} + 1 - F_{T_0,\text{Bell}})(P_{\uparrow,\text{in}}|\delta|^2 + P_{\downarrow,\text{in}}|\zeta|^2) + 2(1 - F_{\Phi,\text{Bell}})(P_{\uparrow,\text{in}}|\zeta|^2 + P_{\downarrow,\text{in}}|\delta|^2)}{P_{S,\text{Bell}}^{\text{model}}}. \quad (5)$$

$\text{Tr}_{12}$ denotes the partial trace over Q1 and Q2. We here estimate errors in the preparation and detection by comparing the experimental results and the model. We first note that there are two kinds of constrains on some of the parameters in Eq. 5 which can be derived from Eqs. 3 and 4. By comparing $P_{\uparrow,\text{out}}^{\text{model}}$ (Eq. 3) and $P_{\uparrow,\text{out}}$, we yield a constraint on the preparation error as $|\delta|^2 + |\zeta|^2 = 0.56$. Similarly, we compare $P_{S,\text{Bell}}^{\text{model}}$ (Eq. 4) to $P_{S,\text{Bell}}$, which is measured as a function of $t_{\text{burst}}$ (data not shown), and obtain another constraint on the detection errors, $F_{S,\text{Bell}} + 1 - F_{T_0,\text{Bell}}$ and $1 - F_{\Phi,\text{Bell}}$ (each value is a function of $|\delta|^2$). By fitting $P_{\uparrow,\text{out}}^{\text{CI,model}}$ to $P_{\uparrow,\text{out}}^{\text{CI}}$ under these two types of constraints, we obtain $P_{\uparrow,\text{out}}^{\text{CI,model}}$ shown by orange squares in Fig. 4 using $F_{S,\text{Bell}} - F_{T_0,\text{Bell}} = 0.38$, $F_{\Phi,\text{Bell}} = 0.96$, $|\delta|^2 = 0.51$ and $|\zeta|^2 = 0.05$. $P_{\uparrow,\text{out}}^{\text{CI,model}}$ and $P_{\uparrow,\text{out}}^{\text{CI}}$ now match each other reasonably well, but there remains a noticeable discrepancy. A possible explanation for this discrepancy is given by considering the estimation error of Q1 readout fidelities, $f_{\text{in},\uparrow}$ and $f_{\text{in},\downarrow}$, which could arise due to a state mapping error in the PSB readout of Q1. Indeed, when we assume $f_{\text{in},\uparrow} = 0.90$ and $f_{\text{in},\downarrow} = 0.93$ (instead of $f_{\uparrow,\text{in}} = 0.96 \pm 0.01$ and $f_{\downarrow,\text{in}} = 0.90 \pm 0.01$), we find $P_{\uparrow,\text{out}}^{\text{CI,model}}$ agrees well with $P_{\uparrow,\text{out}}^{\text{CI}}$ (see blue circles falling on the red line in Fig.4).

Finally, we predict the fidelity of the QT protocol using the obtained errors. We anticipate that our protocol could teleport an arbitrary input state coherently because the 2ns ramp time from (1,0,2) to (2,0,1) is sufficiently shorter than the measured dephasing time of $21 \pm 1$ ns in the device. In this paper, however, we can evaluate only the classical fidelity of the QT protocol because the prepared states of the input qubit are incoherent. We define the classical fidelity to be the probability of finding the spin-up (-down) output qubit for the spin-up (-down) input qubit. With the parameters used for the orange squares and blue circles in Fig. 4, the fidelities are given by $F_\uparrow = P_{\uparrow,\text{out}}^{\text{CI,model}} = 0.95$ for $P_{\uparrow,\text{in}} = 1$ and $F_\downarrow = 1 - P_{\uparrow,\text{out}}^{\text{CI,model}} = 0.86$ for $P_{\uparrow,\text{in}} = 0$ and the averaged fidelity $F_{\text{ave}} = \frac{F_\uparrow + F_\downarrow}{2} = 0.91$. Given our estimations of the fidelities and the parameters, we can now consider the dominant error source in our QT. When we ignore the errors in $\delta$, $\gamma$ and $\zeta$, the fidelities become $F_\uparrow = 0.94$ and $F_\downarrow = 0.94$, indicating the value of $F_\downarrow$ is influenced more significantly by these errors than that of $F_\uparrow$. On the other hand, when we ignore the errors in the Bell measurement, the fidelities become $F_\uparrow = 1.0$ and $F_\downarrow = 0.92$, indicating both of the variations from the original values are comparable. Based on the spin dependence of these fidelities, we attribute the dominant error mechanisms to those related to the spin polarized triplet state $T_+$ leakages. A finite $T_+$ component ($\zeta$) in the prepared singlet pair leads to a bit-flip error only for the spin-down input. In the meantime, erroneous detection of $T_+$ as singlet results in a bit-flip error regardless of the input qubit state. We therefore conclude that the main error source of the classical infidelity is leakage to/from $T_+$ in the preparation and detection process of the singlet state.

The imperfections of preparing and detecting the singlet arise partly from large $\Delta B_x$ and large $\Delta B_z$ between adjacent QDs. The large $\Delta B_x$ induces spin mixing between the ground singlet and $T_+$, leading to the leakage of $T_+$. $\Delta B_z$ between QD1 and QD2 is estimated to be 500 MHz from the ST oscillation in QD1 and QD2 similar to Fig. 2d and it induces the fast relaxation of $|T_0\rangle$ in the Bell measurement decreasing $F_{T_0,\text{Bell}}$. While $F_{T_0,\text{Bell}}$ does not affect the classical fidelities $F_\uparrow$ and $F_\downarrow$, it is important to the quantum mechanical fidelity in coherent teleportation of qubits. Although precise tunings of tunnel couplings can mitigate these problems, they may be avoided by redesigning the micromagnet in future experiments. For example, one can suppress $\Delta B_x$ and $\Delta B_z$ by using a micromagnet in a relatively symmetric geometry with respect to the QD array[32] while maintaining the strong slanting field for EDSR.

In summary, we demonstrate a simple and efficient protocol of the probabilistic QT of a spin qubit in a GaAs TQD device. The ground state initialization of a doubly occupied dot together with a simple pulsed control of detuning allows for the preparation of an entangled state as well as the Bell measurement. The statistics of spin polarization of the output qubit depends on the outcome of the Bell measurement and reproduces that of the input qubit, demonstrating that the spin orientation is teleported from the input qubit to the output qubit. Furthermore, considering our short operation time, we expect that our protocol could teleport an arbitrary input state. We find that the main error source in this protocol is the mixing of the entangled states with $T_+$ substantially due to the large difference of local magnetic fields, which may be improved by optimizing the device design. Our demonstration is among the first demonstrations of teleportation with a single electron spin qubit in semiconductor quantum dots. Our results open a path to demonstrate quantum algorithms with three or more qubits in semiconductor electron spin qubits.

**Acknowledgements:**

Part of this work is financially supported by Core Research for Evolutional Science and Technology (CREST), Japan Science and Technology Agency (JST) (JPMJCR1675 and JPMJCR15N2), the ImPACT Program of Council for Science, Technology and Innovation (Cabinet Office, Government of Japan), a Grant-in-Aid for Scientific Research (JP26220710, JP18H01819, JP19K14640 and JP17K14078), a Grant-in-Aid for JSPS Fellows (JP20J12862), The Murata Science Foundation. Y.K. acknowledges support from Materials Education program for the future leaders in Research, Industry, and Technology (MERIT). T. O. acknowledges support from PRESTO (JPMJPR16N3), JST, Telecom Advanced Technology Research Support Center. AL and ADW greatfully acknowledge financial support from grants DFH/UFA CDFA05-06, DFG TRR160, DFG project 383065199, and BMBF Q.Link.X 16KIS0867.

**Author contributions:**

Y.K., T.N. and S.D.B. conceived and designed the experiment. T.N. and A.N. fabricated the device on the heterostructure grown by A.L. and A.D.W. Y.K. and T.N. performed the measurement and data analysis with the inputs from A.N., J.Y. and K.T. Y.K. and T.N. wrote the manuscript with inputs from other authors. All authors discussed the results and commented on the manuscript. The project was supervised by S.T.

**Competing interests:**

The authors declare that there are no competing interests.

**Data Availability:**

The data that support the findings of this study are available from the corresponding author upon reasonable request.

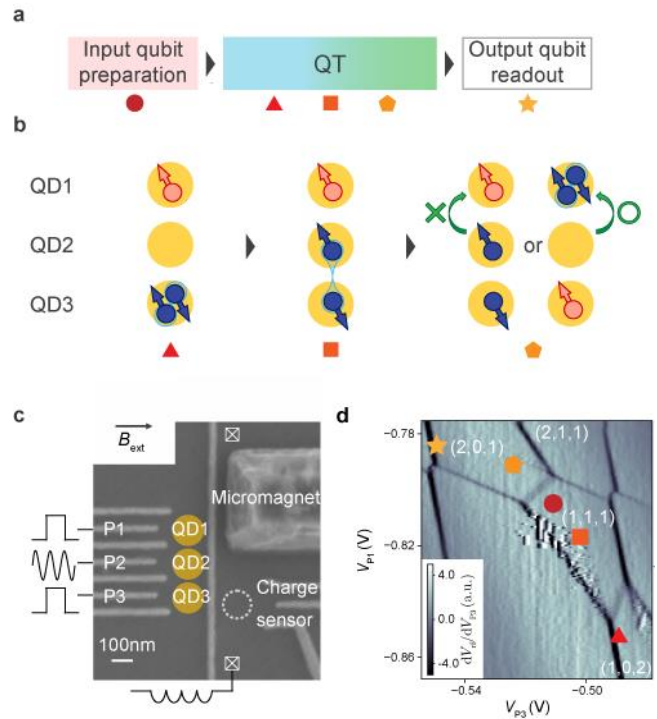

Fig. 1. QT protocol in a TQD. (a) Sequence for the demonstration of QT with markers representing the operation points in the stability diagram in (d). The colored rectangular boxes represent individual steps using the pulse shapes shown in Fig. 2 with corresponding colors. (b) Schematic of our implemantation of the QT protocol. The spin state is teleported from a qubit in QD1 to that in QD3. A spin singlet is prepared in QD3 by adjusting the Fermi level in the reservoir to be between the singlet and triplet levels in QD3. After transferring one of the two electrons in QD3 into QD2, we use Pauli spin blockade (PSB) for the single-shot measurement of the two-spin state in QD1 and QD2 to distinguish whether it is singlet or not. To complete the QT protocol we post-select the single-shot data conditioned on the singlet outcome. (c) Annotated scanning electron micrograph of the TQD similar to the one used for the experiment. A QD charge sensor with rf-reflectometry is used to detect the TQD charge states. (d) Charge stability diagram obtained by sweeping the voltages $V_{\mathrm{PL}}$ and $V_{\mathrm{PR}}$ on the gate electrodes PL and PR. The symbols represent the bias positions used for various operations in the QT experiment (see Fig. 2 and 3).

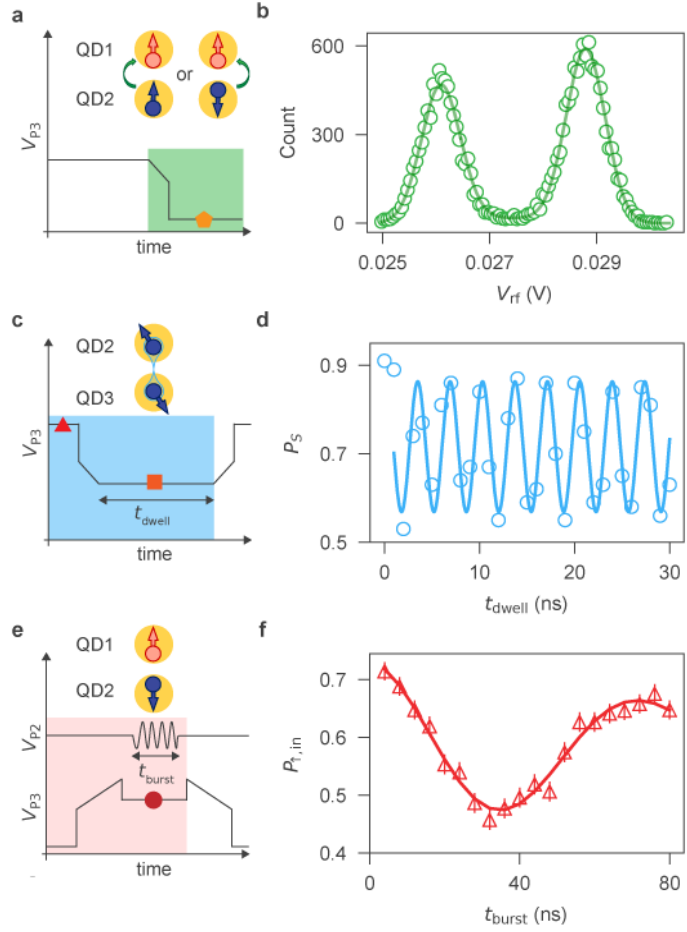

Fig. 2. Ingredients of the QT protocol. (a) Detuning pulse shape represented for $V_{\mathrm{PR}}$ used for the PSB measurement between (1,1,1) and (2,0,1). The detuning is first ramped by rapid adiabatic passage across the (1,1,1) and (2,0,1) resonance and then instantly pulsed to the measurement point (marked by a pentagon) to avoid spin mixing at the anti-crossing of the singlet and the spin-up polarized triplet. (b) Histogram of the single-shot PSB measurement signal $V_{\mathrm{rf}}$. The left population indicates (2,1,1) charge state and the right indicates (1,1,1) charge state. (c) Pulse shape used for preparing a singlet and measuring the singlet-triplet oscillation in QD2 and QD3 while a random spin is left in QD1. A singlet initialized in QD3 at the triangle marker is separated at the square marker and subsequently measured by the ramp similar to the one used in (a). The dwell point (marked by a square) is chosen so that the exchange interaction between Q2 and Q3 is negligible. (d) Singlet-triplet oscillation in QD2 and QD3. The solid line is a fit using the Gaussian decaying envelope. (e) Pulse shape for preparing and measuring the input qubit Q1. To load and measure $|\uparrow_1\Downarrow\rangle$, the ramp is slower than those in (a). A MW burst is applied at the point marked by a circle. (f) Rabi oscillation driven by the resonant SWAP using QD1 and QD2. The MW burst time $t_{\mathrm{burst}}$ is 4, 8, 12, …, 80 ns. Triangles indicate the values of $P_{\uparrow,\mathrm{in}}$ obtained after correcting the readout errors in detecting $|\uparrow_1\Downarrow\rangle$. Error bars represent the standard error. The solid line is a fit using $A_{\mathrm{in}}\cos(2\pi f_{\mathrm{in}} t + \phi_{\mathrm{in}}) e^{-\left(\frac{t}{T_{2,\mathrm{in}}^{\mathrm{Rabi}}}\right)^2} + B_{\mathrm{in}}$ with $A_{\mathrm{in}} = 0.17 \pm 0.02, B_{\mathrm{in}} = 0.57 \pm 0.02, f_{\mathrm{in}} = 13.1 \pm 0.8 \mathrm{MHz}, T_{2,\mathrm{in}}^{\mathrm{Rabi}} = 86 \pm 13 \mathrm{ns}, \phi_{\mathrm{in}} = 0.18 \pm 0.2$. The low visibility is supposed to be due to initialization error of singlet in QD1 because gate voltages for setting the initialization point and dot-lead couplings are not optimal.

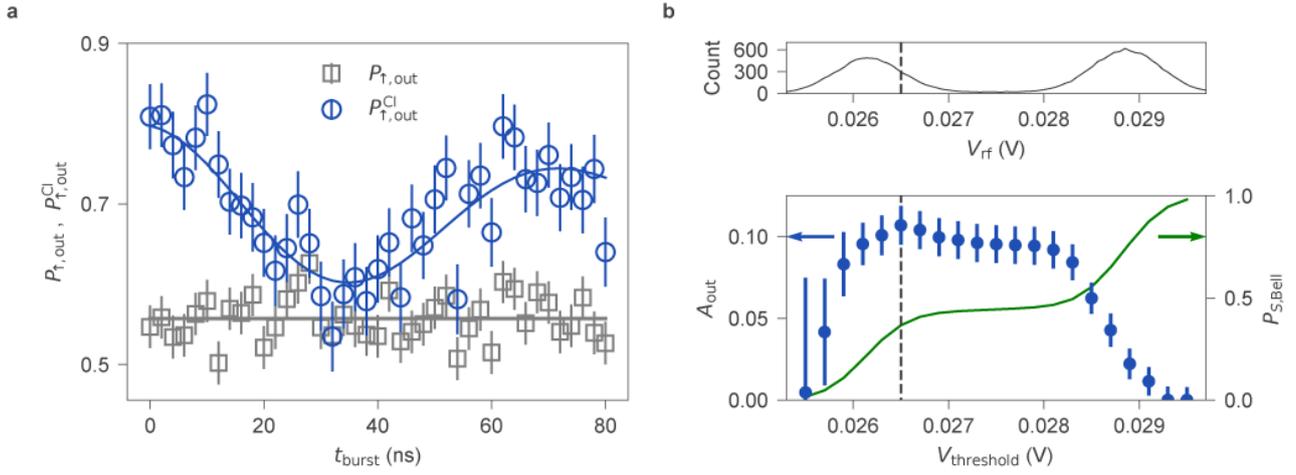

Fig. 3. Demonstration of the QT protocol. (a) Spin-up probability of Q3 obtained as a function of the MW burst time $t_{\text{burst}}$ (0, 2, 4, …, 80 ns) used to vary the spin orientation of Q1. Gray squares are $P_{\uparrow,\text{out}}$ taken regardless of the outcome of the Bell measurement. The gray solid line is a fit to a constant value. Blue circles are $P_{\uparrow,\text{out}}^{\text{CI}}$ extracted from the data conditioned on the singlet outcomes of the Bell measurement. The blue solid line is a fit to

$$A_{\text{out}} \cos(2\pi f_{\text{in}} t + \phi_{\text{in}}) e^{-\left(\frac{t}{T_{2,\text{in}}^{\text{Rabi}}}\right)^2} + B_{\text{out}}$$

with $A_{\text{out}} = 0.11 \pm 0.01$ and $B_{\text{out}} = 0.69 \pm 0.02$, respectively. (b) Dependence of $A_{\text{out}}$ on the threshold voltage $V_{\text{threshold}}$ in the Bell measurement. The threshold voltage used in (a) is shown by a black dashed line. The upper panel shows the histogram of the Bell measurement signal $V_{\text{rf}}$. The lower panel shows $A_{\text{out}}$ and the singlet probability, $P_{S,\text{Bell}}$, as a function of $V_{\text{threshold}}$. $V_{\text{rf}}$ lower (higher) than $V_{\text{threshold}}$ is judged as singlet (triplet). As $V_{\text{threshold}}$ is increased, $A_{\text{out}}$ decreases because the accuracy of the classical information decreases. On the left side of the black dashed line, $A_{\text{out}}$ decreases as the number of singlet outcomes in the Bell measurement decreases and $P_{\uparrow,\text{out}}^{\text{CI}}$ cannot be fitted well.

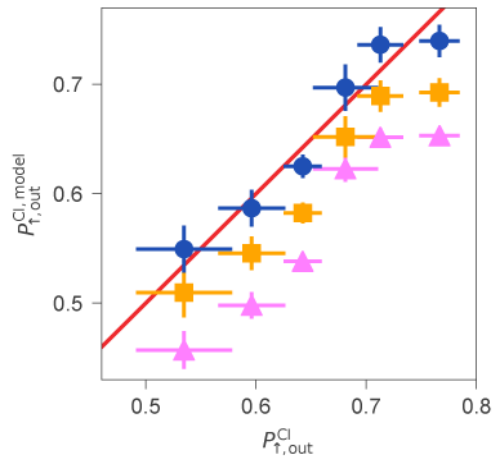

Fig. 4. Correlation between the spin-up probability of the output qubit observed experimentally ($P_{\text{out},\uparrow}^{\text{CI}}$) and expected from the model ($P_{\text{out},\uparrow}^{\text{CI,model}}$). The red line shows the case that $P_{\uparrow,\text{out}}^{\text{CI}}$ is identical to $P_{\uparrow,\text{out}}^{\text{CI,model}}$. $P_{\uparrow,\text{out}}^{\text{CI}}$ is obtained by averaging the scattered data points at each $t_\text{burst}$ in Fig. 3a for clarity. $P_{\text{out},\uparrow}^{\text{CI,model}}$ is calculated by assuming no errors (pink triangles), finite errors in preparation and detection of singlet (orange squares), and inaccurate estimation of the input qubit readout fidelities in addition to the preparation and detection errors (blue circles), respectively. In the calculation we use estimated values of parameters of $F_{S,\text{Bell}} - F_{T_0,\text{Bell}} = 0.38$, $F_{\Phi,\text{Bell}} = 0.96$, $|\delta|^2 = 0.51$ and $|\zeta|^2 = 0.05$ (for orange squares and blue circles), and $f_{\text{in},\uparrow}$ ($f_{\text{in},\downarrow}$) $= 0.90$ (0.93) (for blue circles).

**Supplementary material for Probabilistic teleportation of a quantum dot spin qubit**

Y. Kojima *et al*.

## I. Inter-dot tunnel couplings between adjacent QDs

It is assumed in QT that the output qubit is not directly coupled to the input qubit. In the TQD, there may be a direct exchange interaction between spins in QD1 and QD3 when large tunnel couplings between adjacent QDs are simultaneously turned on. To estimate the nearest-neighbor tunnel coupling, we measure the energy difference $\Omega$ between the two eigenstates in the $ST_0$ subspace, using a resonant SWAP technique. First we load a singlet in (2,0,1) ((1,0,2)) from the reservoir and ramp the detuning around the zero-detuning point (1,1,1) (in Fig. 1c) using slow adiabatic passage to initialize into $|\uparrow\downarrow\rangle$ ($|\downarrow\uparrow\rangle$)[12]. Then, we instantaneously pulse to the operation point and modulate $J$ by applying MW to the PC gate to swap the two spins resonantly. The resonance frequency is equivalent to $\Omega$. Finally, we adopt the reversed gate pulse to distinguish $|\uparrow\downarrow\rangle$ and $|\downarrow\uparrow\rangle$ in the similar technique to PSB and measure the spin flip probability. The results of the detuning dependence of the probabilities are shown in Fig. S1.

To extract the $\epsilon$ dependence of $\Omega$, we fit the data in Fig. S1. We get the value of $\Omega$ by taking the MW frequency where spin non-flip probability is minimum at a fixed detuning. We use $\Omega = \sqrt{J^2 + \Delta B^2}$, where $J = \frac{1}{2}\left[-A \times (\epsilon - \epsilon_0) + \sqrt{A^2(\epsilon - \epsilon_0)^2 + 8t_c^2}\right]$ with $\epsilon_0 = -21$ mV for QD1-QD2, as a fitting function. The data for QD1-QD2 can be fitted well to yield $A = 445 \pm 31$ MHz/mV, $t_c = 354 \pm 6$ MHz and $\Delta B_{\text{QD1QD2}} = 488 \pm 2$ MHz, respectively. The deviation from $\Delta B_{\text{QD1QD2}} = 500$ MHz extracted from ST oscillation (in main text) is due to the fluctuation of nuclear spin field. We calculate $J_{\text{QD1QD2}} = 27$ MHz at $\epsilon = 0$ mV. In contrast, the data of QD2-QD3 is difficult to fit because the dip of spin non-flip probability is faint and we cannot precisely extract the value of $\Omega$. However, the faintness and the difficulty in performing resonant SWAP empirically implies that the inter-dot tunnel coupling is much weaker than in QD1-QD2 and $J_{\text{QD2QD3}}$ can be ignored in the (1,1,1) region. Considering the operation time required for our QT is a few nanoseconds, which is much faster than the inverse of the adjacent exchange couplings, we conclude that direct interaction between QD1 and QD3 has no effect on our results.

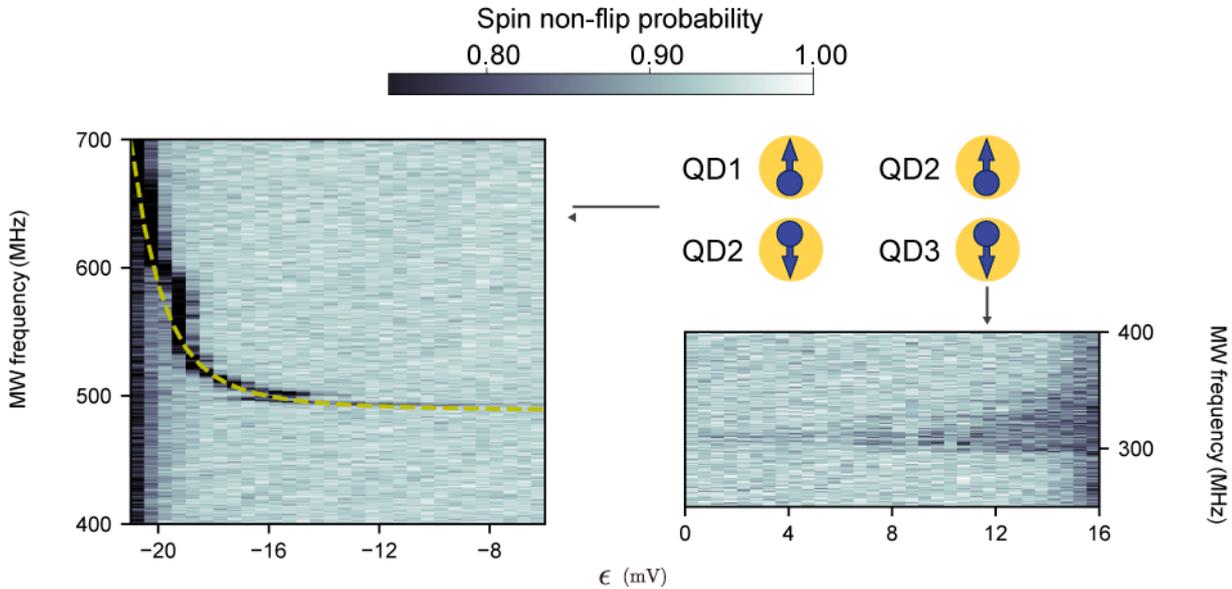

Fig. S1. (a), (b) Spin non-flip probability dependence of detuning $\epsilon$ and the resonant SWAP microwave frequency. The yellow dashed line is the fitting result of $\Omega$. The MW burst time is 700 ns which is long enough to see the spectrum.

## II. Estimation of the readout fidelity of input qubit

We can calculate the readout fidelities of input qubit from the signal separation between two charge states and the noise broadening of the charge sensor. Figure S2 is the histogram of single-shot readout result of PSB. It shows a bimodal distribution having a singlet peak at $V_S$ and a triplet peak at $V_T$. The threshold voltage is chosen to maximize the averaged charge state readout fidelities. When the input spin state is spin-up (spin-down), the single-shot outcome of PSB is singlet (triplet). When we consider the model taking into account the relaxation of triplet states, we obtain $f_{\uparrow,\text{in}} = 0.96$ and $f_{\downarrow,\text{in}} = 0.90$.

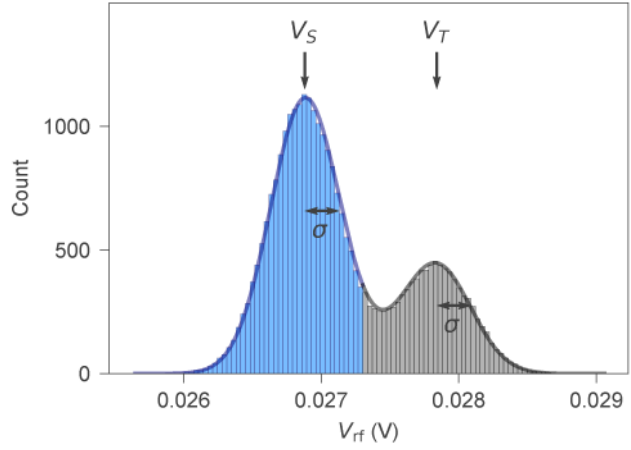

Fig. S2. The histogram of the PSB measurement results in the readout of Q1. The left population peak is for a singlet outcome (Q1 is spin-up) and the right is for triplet (Q1 is spin-down). The solid line is the fitting result using the model taking into account the relaxation of triplet states. We get $V_S = 0.02688$, $V_T = 0.02784$, $\sigma = 0.0002508$.

## III. Estimation of the readout fidelity of output qubit

We use the spin-selective tunneling to the reservoir for readout of output qubit. In this readout method we judge the spin state by whether or not to detect the blip signal caused by electron tunneling events in the measurement time. To estimate the readout fidelity of the output qubit we use the model of Keith et al[30]. The error of the readout is decomposed into two factors. One is the error in the state-to-charge conversion (STC) and the other is the error in electrical detection of the blip signal. The readout fidelity $f_{\text{out},\uparrow}$ and $f_{\text{out},\downarrow}$ can be described by

$$f_{\text{out},\uparrow} = F_{\text{STC}}^{\uparrow} F_E^{\uparrow} + (1 - F_{\text{STC}}^{\uparrow})(1 - F_E^{\downarrow})$$
$$f_{\text{out},\downarrow} = F_{\text{STC}}^{\downarrow} F_E^{\downarrow} + (1 - F_{\text{STC}}^{\downarrow})(1 - F_E^{\uparrow})$$

Here we denote by $F_{\text{STC}}^{\downarrow}$ ($F_{\text{STC}}^{\uparrow}$) the probabilities that $\downarrow$ ($\uparrow$) does (not) tunnel to the reservoir and denote by $F_E^{\downarrow}$ ($F_E^{\uparrow}$) the probabilities that we do (not) detect a blip signal for where there is a (no) tunneling event.

To estimate $F_{STC}^{\uparrow}$ and $F_{STC}^{\downarrow}$, we examine the tunnel rate $t_{OUT}^{\uparrow(\downarrow)}$ from QD3 to the reservoir. We manipulate the spin in QD3 by EDSR with a frequency-chirped MW and realize the spin-selective tunneling between QD3 and the reservoir. The distribution of the electron tunneling time gives the tunnel rate of each state. In our measurement condition, we find $t_{OUT}^{\uparrow} = 1055$ μs and $t_{OUT}^{\downarrow} = 263.9$ μs from fitting of the histogram of detection time of each spin state (Fig. S3). With $T_1 = 1.57$ ms found in a different cool down[15] and measurement time length 200 μs, we get

$$F_{STC}^{\uparrow} = 0.827 \text{ and}$$
$$F_{STC}^{\downarrow} = 0.511.$$

To estimate $F_E^{\uparrow}$ and $F_E^{\downarrow}$, we examine signal-to-noise ratio of the charge sensor. Assuming that the noise distribution of the charge sensor signal of each state is Gaussian distribution, we get $V_{\uparrow} = 0$ V, $V_{\downarrow} = 0.003767$ V, $\sigma_{\uparrow} = 0.0003204$ V and $\sigma_{\downarrow} = 0.0003567$ V (Fig. S4), where $V_{\uparrow}$ ($V_{\downarrow}$) are the mean values and $\sigma_{\uparrow}$ ($\sigma_{\downarrow}$) are the standard deviations of the sensor signal of each state. With the sampling rate 1 μs and $t_{IN}^{\uparrow} = 15.84$ μs (Fig. S3), we get

$$F_E^{\uparrow} = 1.0 \text{ and}$$
$$F_E^{\downarrow} = 0.982.$$

Therefore, the readout fidelities $f_{out,\uparrow}$ and $f_{out,\downarrow}$ become

$$f_{out,\uparrow} = 0.83 \text{ and}$$
$$f_{out,\downarrow} = 0.50.$$

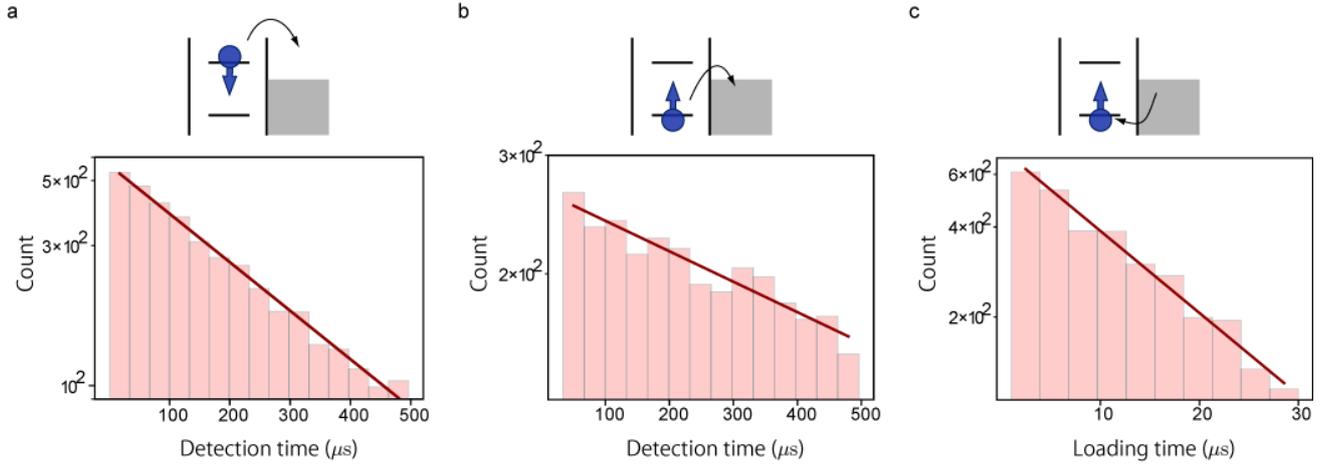

Fig. S3. (a), (b), Histograms of detection time of the charge sensor signal jump. After we apply a frequency-chirped MW to the center horizontal gate in the Coulomb blockade region of (2,0,1), we ramp to the point close to the charge transition line between (2,0,1) and (2,0,0) and monitor the signal of the charge sensor. (a)((b)) shows the case where the spin down (up) is prepared by chirping the MW frequency around the resonance frequency before it is brought to the readout position and tunnels out. In other words, the slope of (a) gives the tunnel out time of spin-down electrons, $t_{OUT}^{\downarrow}$, and the slope of (b) gives the tunnel out time of spin-up electrons, $t_{OUT}^{\uparrow}$. We obtain $t_{OUT}^{\downarrow} = 263.9$ μs and $t_{OUT}^{\uparrow} = 1055$ μs. (c) Histogram of loading time of an electron into QD3. The slope gives the tunnel in time of spin-up electrons in QD3, $t_{IN}^{\uparrow}$. We obtain $t_{IN}^{\uparrow} = 15.84$ μs.

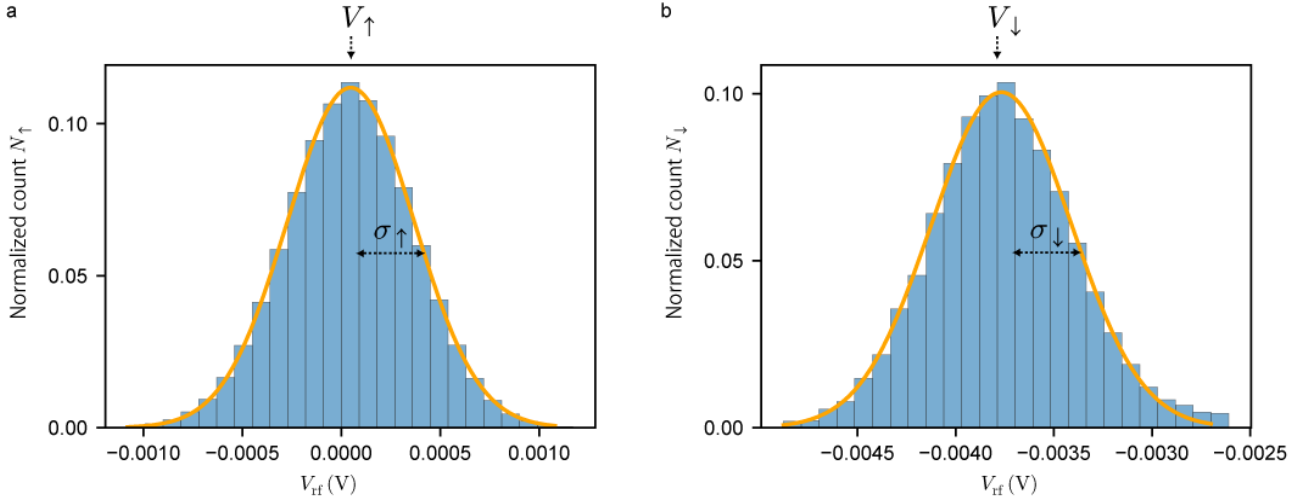

Fig. S4. Histograms of the sensor fluctuation for the (2,0,1) charge configuration (a) and the (2,0,0) configuration (b). The orange lines are the results of gaussian distribution fitting $N_i = a_i \times \exp\left[-\frac{(V_{\rm rf}-V_i)^2}{2\sigma_i^2}\right]$. We obtain $V_\uparrow = 0.000$ V, $V_\downarrow = -0.003767$ V, $\sigma_\uparrow = 0.0003204$ V and $\sigma_\downarrow = 0.0003567$ V.

## IV. Complementary explanation for $P_{\uparrow,\rm out}^{\rm CI,model}$

$P_{\uparrow,\rm out}^{\rm CI,model}$ is written by $\frac{(F_{S,\rm Bell}+1-F_{T_0,\rm Bell})(P_{\uparrow,\rm in}|\delta|^2+P_{\downarrow,\rm in}|\zeta|^2)+2(1-F_{\Phi,\rm Bell})(P_{\uparrow,\rm in}|\zeta|^2+P_{\downarrow,\rm in}|\delta|^2)}{P_{S,\rm Bell}^{\rm model}}$. The prepared state in Q2 and Q3 is given by $\gamma|\uparrow_2\downarrow_3\rangle + \delta|\downarrow_2\uparrow_3\rangle + \zeta|\uparrow_2\uparrow_3\rangle$ and the terms contributing to obtain spin-up in Q3 are $\delta|\downarrow_2\uparrow_3\rangle$ and $\zeta|\uparrow_2\uparrow_3\rangle$. In our QT protocol, when the parity of the Q2 and Q3 states prepared is identical to that of the Bell measurement outcome on Q1 and Q2, the orientation of Q1 is teleported to that of Q3. Otherwise, a bit-flip occurs on Q3. Combinations for the parities of the prepared states and the projected states are divided to four patterns as summarized in the following table. For example, with $|\downarrow_2\uparrow_3\rangle$, Q1 is teleported to Q3 when Q1 and Q2 is projected to $|S_{12}\rangle$ or $|T_{0,12}\rangle$, whereas it will result in a bit flip error when Q1 and Q2 is projected to $|\Phi_{12}\rangle$. We calculate the probability that Q3 is spin-up and a singlet is detected in the Bell measurement, $P(\uparrow_3 \cap S_{12})$, as the product of the probability of the prepared state in Q2 and Q3, the spin-up probability of Q3 for the projected state in the Bell measurement and the probability of judging the case as singlet in the Bell measurement. $P_{\uparrow,\rm out}^{\rm CI,model}$ is given by the sum of $P(\uparrow_3 \,|S_{12}) = \frac{P(\uparrow_3 \cap S_{12})}{P(S_{12})}$ for each case.

Table S1. The combinations for prepared state in Q2 and Q3 and projected state in Q1 and Q2, and their probabilities.

| The prepared state in Q2 and Q3 (its probability) | The projected state in the Bell measurement (the spin-up probability of Q3) | The probability of being judged as singlet in the Bell measurement |
|---|---|---|
| $|\downarrow_2\uparrow_3\rangle$ ($|\delta|^2$) | $|S_{12}\rangle, |T_{0,12}\rangle$ ($P_{\uparrow,\rm in}$) | $F_{S,\rm Bell}$, $1 - F_{T_0,\rm Bell}$ |
| $|\downarrow_2\uparrow_3\rangle$ ($|\delta|^2$) | $|\Phi_{12}\rangle$ ($P_{\downarrow,\rm in}$) | $1 - F_{\Phi,\rm Bell}$ |
| $|\uparrow_2\uparrow_3\rangle$ ($|\zeta|^2$) | $|S_{12}\rangle, |T_0\rangle$ ($P_{\downarrow,\rm in}$) | $F_{S,\rm Bell}$, $1 - F_{T_0,\rm Bell}$ |
| $|\uparrow\uparrow_{23}\rangle$ ($|\zeta|^2$) | $|\Phi_{12}\rangle$ ($P_{\uparrow,\rm in}$) | $1 - F_{\Phi,\rm Bell}$ |